\documentclass[aip,jcp,amsmath,amssymb,reprint]{revtex4-1}
\usepackage[english]{babel}
\usepackage[utf8]{inputenc}
\usepackage{amsmath}
\usepackage{amssymb}
\usepackage{siunitx}
\usepackage[textsize=small]{todonotes} 
\usepackage{subcaption}
\usepackage{fancyhdr}
\newcommand{\kb}{k_\textnormal{B}}

\renewcommand{\vec}[1]{\mathbf{#1}}
\fancypagestyle{legalnotice}{%
  \fancyhead[L]{}%
  \fancyhead[R]{}%
  \fancyfoot[C]{}%
  \fancyfoot[R]{\footnotesize Page \thepage\ of \pageref{LastPage}}%
  \fancyfoot[L]{\footnotesize The following article has been submitted to/accepted by the Journal of Chemical Physics.\\After it is published, it will be found at \url{https://aip.scitation.org/journal/jcp}.}%
}

\begin{document}
\title{Modeling the Current Modulation of Bundled DNA Structures in Nanopores}

\author{Kai Szuttor}
\email[]{kai@icp.uni-stuttgart.de}
\author{Florian Weik}
\author{Jean-No\"el Grad}
\author{Christian Holm}
\email[]{holm@icp.uni-stuttgart.de}
\affiliation{Institute for Computational Physics, Universit\"at Stuttgart, Allmandring 3, D-70569 Stuttgart, Germany}

\date{\today}
\begin{abstract}
  We investigate the salt-dependent current modulation of bundled DNA
  nanostructures in a nanopore. To this end, we developed four
  simulation models for a 2x2 origami structure with increasing level
  of detail ranging from the mean-field level to an all-atom representation
  of the DNA structure. We observe a consistent pore conductivity as a
  function of salt concentration for all four models.  However, a
  comparison of our data to recent experimental investigations on
  similar systems displays significant deviations. We discuss possible
  reasons for the discrepancies and propose extensions to our models
  for future investigations.
\end{abstract}
\maketitle
\thispagestyle{legalnotice}
\pagestyle{legalnotice}

\section{Introduction}
The field of nanopore based molecule detection and analysis has shown growing
interest in the soft matter scientific community in the last years. The basic
idea dates back to the late 1940s when Wallace Coulter \cite{coulter49a}
invented a device to count red blood cells in a setup consisting of two
electrolyte reservoirs in an electric field. A small orifice connecting the two
reservoirs lead to a finite conductivity. For the pure salt solution the
conductivity was constant, whereas spikes in the current have been observed if a
blood cell traverses the orifice which made it possible to count the cells. More
refined setups allowed detecting smaller and smaller analytes
\cite{deblois70a, kasianowicz96a}. Leveraging this simple principle, nanopores
are nowadays capable of detecting single DNA molecules, and even a distinction
between nucleotides is possible\cite{cherf12a,manrao12a}.

For dsDNA, it is well-known that the current modulation through the pore depends
on the salt concentration of the buffer solution \cite{smeets06a}. For salt
concentrations below a certain crossover concentration $\tilde{c}$, the current
increases due to the presence of the DNA whereas for higher salt concentrations
the current decreases. So far, experimental results for the current modulation
have been reproduced quantitatively with molecular-dynamics simulations on
different levels of detail: with an all-atom model \cite{kesselheim14a}, with a
coarse-grained model \cite{weik16a} and with a mean-field description for the
dsDNA and the electrolyte \cite{weik19b}. These studies revealed that the
current blockage in the pore is not caused by a reduction of the number of
charge carriers (up to \SI{1.2}{\mol\per\litre}) in the pore but by the
enhanced local friction between ions and DNA. Similar to the mean-field
and all-atom models in Refs.~\onlinecite{weik19b, kesselheim14a}, Van Dorp
\textit{et al.} had developed a mean-field model and Luan \textit{et al.} 
an all-atom model to study the electrophoretic forces on the DNA in
Refs.~\onlinecite{vandorp09a, luan08a}. 

In recent experiments the self-assembly of scaffolded three-dimensional DNA
origami structures has been investigated \cite{castro11a,han11a,yoo13a}. In various
publications, such DNA origami structures have been used to alter the analyte
detection properties of nanopores in experiments \cite{bell12a, langecker12a,
wei12b, hernandez13a, hernandez14a, goepfrich16a} and simulation studies
\cite{li15b, goepfrich16a, barati17a}.

Wang \textit{et al}. investigated the current signal change caused by the translocation
of DNA origami bundles with 4 and 16 parallel helices\cite{wang19a}. Their
findings show that the salt dependent current modulation significantly changed
compared to the results for the translocation of a single dsDNA molecule
\cite{smeets06a}. The crossover salt concentration where the current modulation
by the analyte in the pore vanishes drops to much smaller values. Furthermore,
this effect shows a non-monotonic dependency on the size of the origami
structure whose origin remains unclear.

In Ref.~\onlinecite{weik19b}, the results for the current modulation
for dsDNA in a nanopore can successfully be modelled on the all-atom,
the coarse-grained and even the mean-field scale. Thus, the natural
question arises if these models are trivially transferable to multiple
helices, \textit{i.e.} can the overall interaction between ions and
DNA be predominantly described by a linear superposition of the
interaction with a single helix. To find an answer for this issue, we have
therefore investigated the current modulation of bundled DNA origami
structures in an infinite cylindrical pore on different levels of
detail, ranging from all-atom to mean-field. Remarkably, all four
models' results for the pore conductivity are consistent with each
other within statistical errors.

The article is structured as follows: in Sec.~\ref{sec:models}, we first
describe the four different simulation models, in Sec.~\ref{sec:results} we
present our simulation results and compare them to experimental results in
Sec.~\ref{sec:exp}. Our article ends with some conclusions and an
outlook for further studies.

\section{DNA models}\label{sec:models} In the following we describe the DNA
modeling approaches in more detail. All our models have in common that they
only contain the pore segment (and no reservoir) which enables us to reduce the
computational effort by applying periodic boundary conditions along the symmetry
axis of the pore. Since experimental results for the current modulation of dsDNA
in solid-state nanopores \cite{smeets06a} and in glass nanocapillaries
\cite{steinbock12a} are very similar, we have chosen to take advantage of the
more symmetric cylindrical geometry. 

\subsection{All-atom DNA origami}\label{sec:origami_model} The basis of our
all-atom model is a structure file of the DNA origami provided to us
by the Keyser group which did the experimental work motivating this
study \cite{wang19a}. We extracted a periodically recurring segment of
the origami and placed it in a rectangular simulation box with
periodic boundary conditions to create an infinite origami that
reproduced the shape of the original structure, using bonded
interactions across the unit cell.  We added a cylindrical pore wall
built up of atoms with a purely repulsive interaction potential.  The
dsDNA origami's phosphorus atoms as well as the pore atoms have been
fixed in space via a harmonic potential. We solvated the molecule in
water and compensated the net charge by an appropriate amount of
potassium counter-ions. To simulate different bulk salt concentrations
we exchanged a varying number of water molecules with potassium
chloride ion pairs.  For the dsDNA molecule we employed the AMBER03
force field~\cite{duan03a}, for the salt ions the force field by
Smith~et~al.\cite{smith94b} and for the water the SPC/E water
model~\cite{berendsen87a}. This combination of force fields has
previously been used in similar simulation setups and has shown to
reproduce experimental ion conductivity very
well~\cite{kesselheim14a}.  The bulk salt concentrations have been
estimated a posteriori from the charge density profiles. The all-atom 
simulations have been performed with the molecular dynamics software
GROMACS version 2020.3 \cite{abraham15a}.

\begin{figure}[ht]
	\begin{subfigure}{0.49\linewidth}
		\centering
		\includegraphics[height=5cm]{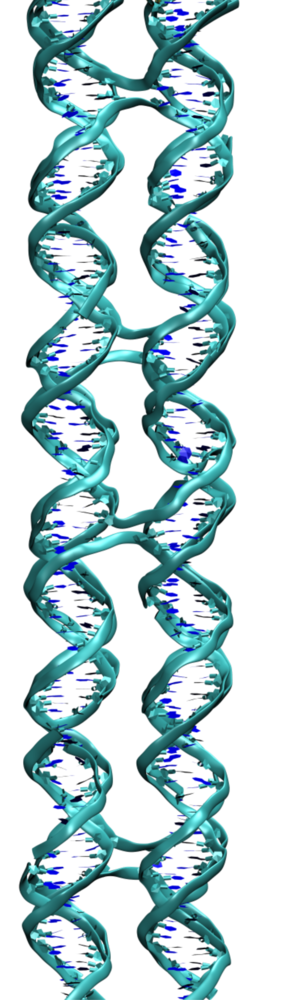}
	\end{subfigure}
	\hfill
	\begin{subfigure}{0.49\linewidth}
		\centering
		\includegraphics[width=\linewidth]{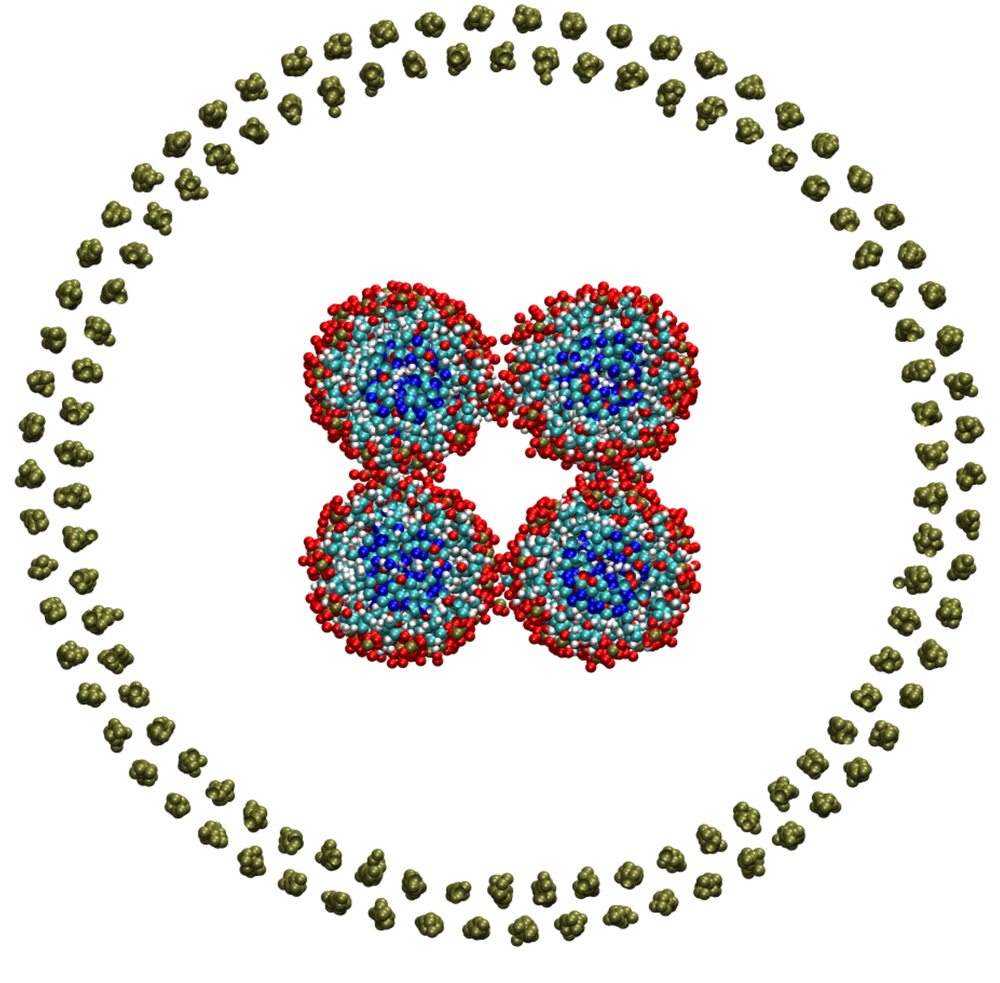}
	\end{subfigure}
	\caption{The side view (left) and cross-section (right) of the
          all-atom origami molecule section in a pore. For clarity the
          water molecules have been excluded.}
	\label{fig:aaorigami}
\end{figure}

Details of the procedure of setting up the periodic chunk of the origami
molecule can be found in the supporting information.  In the visualization of
the simulated molecule in Fig.~\ref{fig:aaorigami} the inter-helix staple
strands that stabilize the origami structure are visible.

\subsection{All-atom quad}\label{sec:quad_model} 
\begin{figure}[ht]
	\begin{subfigure}{0.49\linewidth}
		\centering
		\includegraphics[height=5cm]{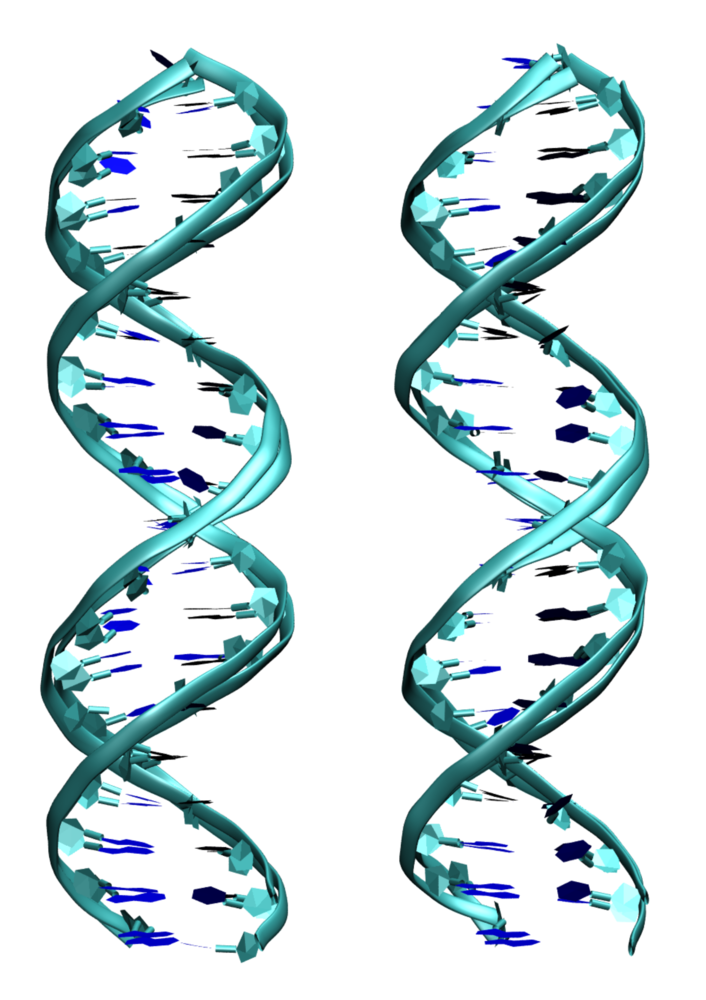}
	\end{subfigure}
	\hfill
	\begin{subfigure}{0.49\linewidth}
		\centering
		\includegraphics[width=\linewidth]{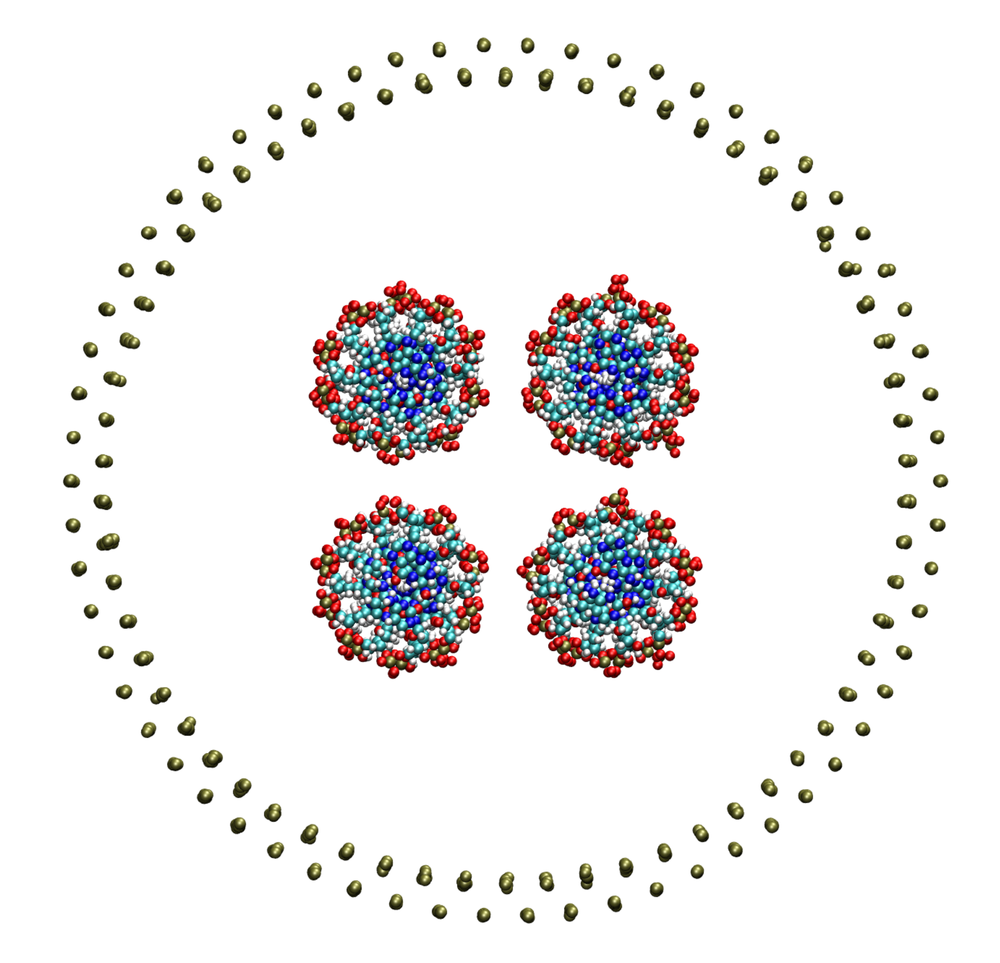}
	\end{subfigure}
	\caption{Side view (left) and cross-section (right) of the
          simulated all-atom quad molecule in a pore.}
	\label{fig:aaquad}
\end{figure}

This model is constructed by
placing four parallel strands of the dsDNA described in
Ref.~\onlinecite{kesselheim14a} parallel to each other. The main difference to
the model in the previous section is the missing interconnecting staple strands
(Fig.~\ref{fig:aaquad}). Each of the four helices consists of 20 CG base pairs
which corresponds to two full turns of the helix. All other simulation
parameters have been adopted from the model above.

\subsection{Coarse-grained DNA origami}\label{sec:cg_model} 
\begin{figure}[ht]
	\begin{minipage}{0.49\linewidth}
		\centering
		\includegraphics[width=\linewidth]{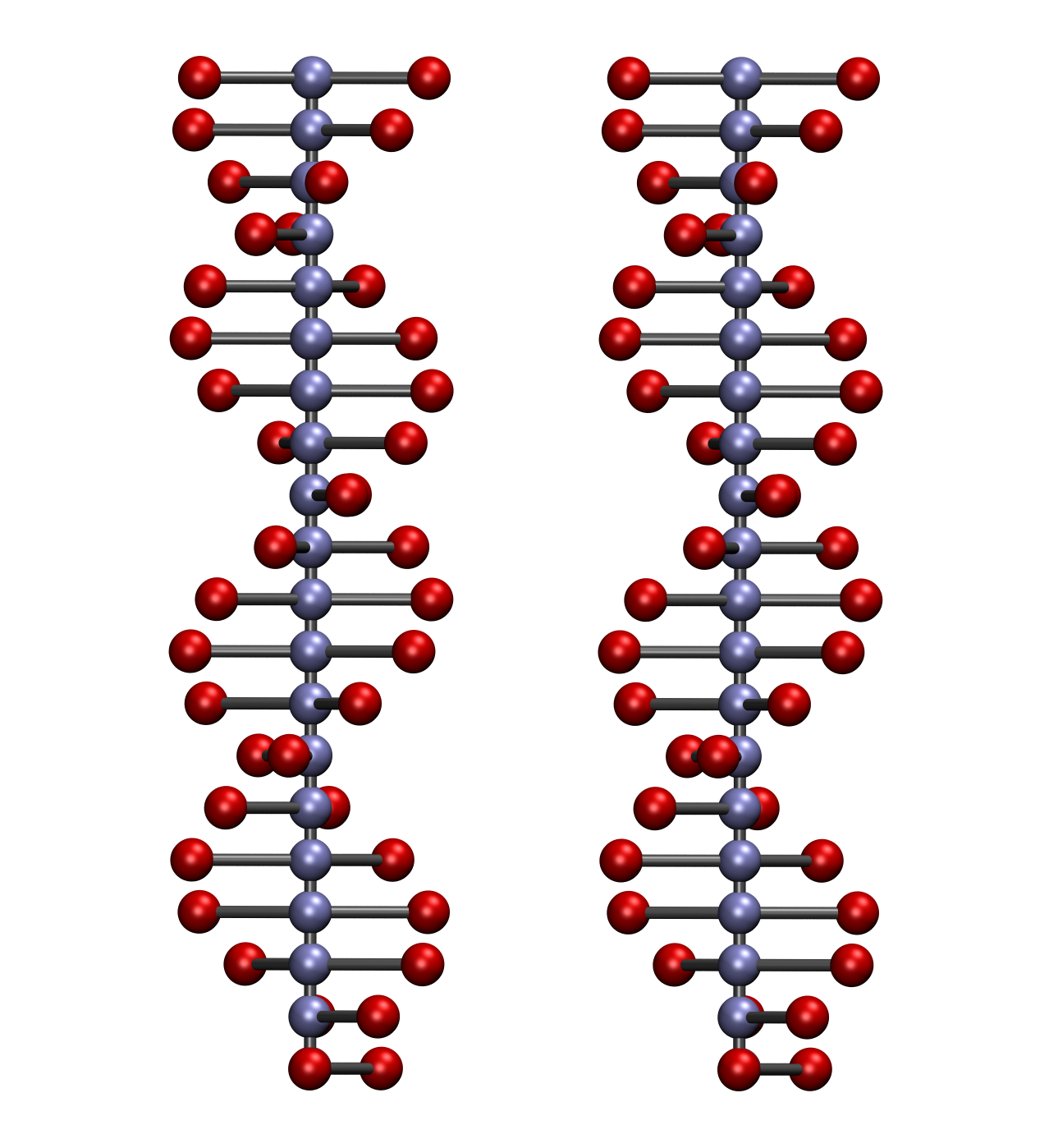}
	\end{minipage}
	\hfill
	\begin{minipage}{0.49\linewidth}
		\centering
		\includegraphics[width=\linewidth]{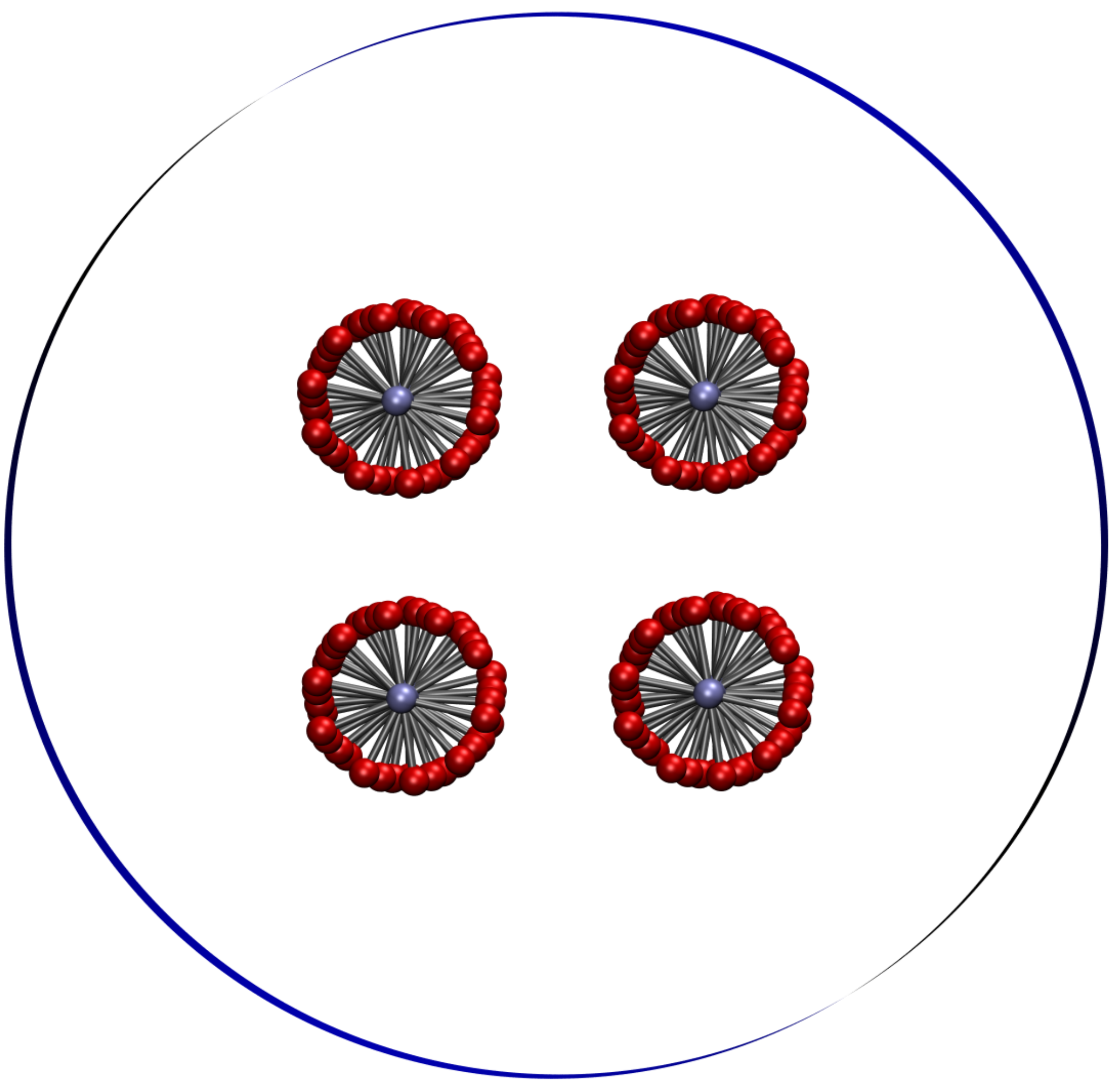}
	\end{minipage}
	\caption{The side view (left) of the coarse-grained quad molecule as
		developed in Ref.~\onlinecite{rau17a} and the cross-section (right) of the
		simulated pore system.}
	\label{fig:cgdna}
\end{figure}

We followed the
modelling strategy for a coarse-grained model of a dsDNA molecule as published
in Ref.~\onlinecite{rau17a}. In order to extend the model to reproduce the
geometry of the origami molecule we created a fixed arrangement of four parallel
dsDNA strands (\textit{cf.} Fig.~\ref{fig:cgdna}). Simulations have been
performed with the release 4.0 of the MD software ESPResSo \cite{weik19a}.

\begin{figure}[ht]
	\centering
	\includegraphics[width=\linewidth]{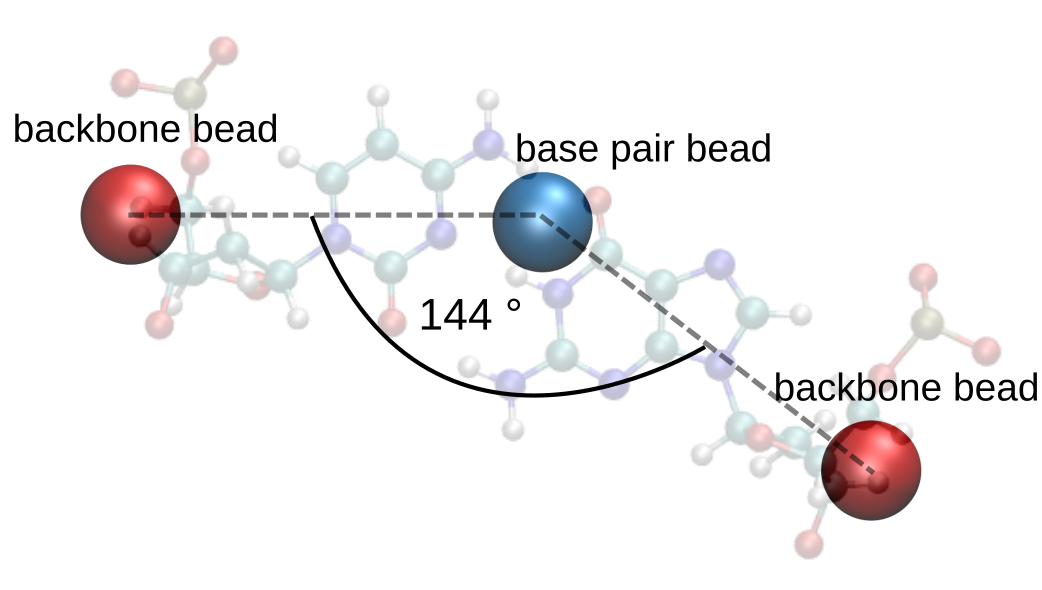}
	\caption{Top view of a single coarse-grained base pair. In the background
	the respective atomic visualization is shown for comparison.
	Note, that the beads are not drawn to scale for clarity.}
	\label{fig:coarse_grained}
\end{figure}

The segments of the coarse-grained dsDNA consist of a rigid arrangement of three
beads (two for the backbone and one for the base pair). A scheme for a single
base pair is shown in Fig.~\ref{fig:coarse_grained}. Hydrodynamic interactions
are included by coupling molecular dynamics with a lattice Boltzmann
hydrodynamics solver. Using the point coupling scheme \cite{ahlrichs99a,
duenweg09a} to exchange momentum between particles and the lattice Boltzmann
fluid, a nonphysical fluid flow has been observed along the grooves of the helix
\cite{weik16a} which is suppressed with additional beads that only interact with
the fluid and are part of the rigid body of each segment. To match the distance
dependent ion mobility in the vicinity of the DNA observed in all-atom
simulations \cite{kesselheim14a}, a frictional coupling force $\vec{F}_{ij}$
between the backbone particles and the ions is added:
\begin{align}
	\vec{F}_{ij} = \begin{cases}
		-\zeta \left(1-\frac{r_{ij}}{r_\mathrm{c}}\right)^2\vec{v}_{ij}, & r_{ij} \leq r_\mathrm{c}\\
		0, & \mathrm{else},
	\end{cases}
\end{align}
where $\zeta$ is a numerical constant, $r_{ij}$ is the inter-particle distance
between DNA bead and ion, $r_\mathrm{c}$ is the cutoff radius up to which the
frictional interaction is enabled, $\vec{v}_{ij}$ is the relative velocity. The
two free parameters $\zeta$ and $r_\mathrm{c}$ are tuned to match the ion
velocity profile of the respective all-atom simulation for a single salt
concentration. In addition to the frictional force, a random force according to
the fluctuation-dissipation theorem is applied. Details of the model can be
found in Ref.~\onlinecite{rau17a}, exact values for all parameters are listed in the supporting
information.

\subsection{Mean-field model}\label{sec:mean_field_model} Based on the geometric
arrangement of the coarse-grained model in the previous section, the DNA strands
in the mean-field model are replaced by charged cylinders that carry the
respective surface charge density of a dsDNA molecule ($\sigma_{\text{DNA}}
\approx\SI{-2}{\elementarycharge}/ \SI{2.35}{\nano\metre\squared}\approx
\SI{-0.136}{\coulomb\per\meter\squared}$). It is based on solving the modified
electrokinetic equations for the bundle of charged cylinders representing the
dsDNA structure within an uncharged cylinder representing the nanopore following
the approach in Ref.~\onlinecite{weik19b}. The Nernst-Planck equation is
modified to incorporate the friction between ions and dsDNA molecules that has
previously been found to be crucial for coarsened dsDNA models in order to
reproduce experimental data on current modulation in nanopores~\cite{weik16a}.
Thus, the algebraic equation for the current density of species $i \in
\{\mathrm{K},\mathrm{Cl}\}$ along the symmetry axis reads:
\begin{align}
	j_i^z = \Biggl[\underbrace{\vphantom{\left(\frac{j_i^z}{c_i}\right)}u^z}_{\text{advection}} + 
	\mu_i\Biggl(\underbrace{\vphantom{\left(\frac{j_i^z}{c_i}\right)}e z_i E^z}_{\text{external field}} \underbrace{\vphantom{\left(\frac{j_i^z}{c_i}\right)}- \alpha \omega \frac{j_i^z}{c_i}}_{\text{friction}}\Biggr)\Biggr]c_i,
\end{align}
where $c_i$ is the concentration, $u^z$ is the fluid velocity, $\mu_i$ the ion mobility, $e$ the
elementary charge, $z_i$ the valency, $E^z$ the electric field strength
along the symmetry axis, $\alpha=\SI{15e-12}{\kilo\gram\per\second}$ a numerical constant (controlling the
amount of frictional force) and $\omega$ is a position dependent weight
function for the frictional force (\textit{cf.} Ref.~\onlinecite{weik19b} for details).

The advective motion of ions is taken into account by means of the Stokes' equation
and electrostatic interactions between ions are considered following the
Poisson-Boltzmann theory. Due to the friction between ions and dsDNA this
results in a coupling between electrostatic and advective forces on the ions.
The model's inherent symmetry reduces the problem domain to two
dimensions. We solved the electrokinetic equations with the finite-element
method using the commercial software package COMSOL
Multiphysics\textsuperscript{\textregistered} version 5.4.

\begin{figure}[htbp]
    \begin{tikzpicture}[scale=0.4]
        \draw (0, 0) circle (5.0);
		\draw[blue] (-1.5, -1.5) circle (1.1);
		\coordinate (x) at (-2.6, -2.7);
		\coordinate (y) at (-0.4, -2.7);
		\draw[thick, <->] (x) -- (y) node[below, pos=0.8] {$d_{\mathrm{DNA}}=\SI{2.2}{\nano\meter}$};
        \draw[blue] (-1.5, 1.5) circle (1.1);
        \draw[blue] (1.5, 1.5) circle (1.1);
        \draw[blue] (1.5, -1.5) circle (1.1);

        \coordinate (x) at (-1.5, 1.5);
        \coordinate (y) at (-1.5, -1.5);
        \draw [thick, <->] (x) -- (y) node[right, pos=0.5] {$d_{\mathrm{origami}}=\SI{2.8}{\nano\meter}$};
        \coordinate (x) at (-1.5, -1.5);
        \coordinate (y) at (1.5, -1.5);

        \draw [dashed] (-5.0, 0.0) -- (-5.0, -6.0);
        \draw [dashed] (5.0, 0.0) -- (5.0, -6.0);
        \coordinate (x) at (-5.0, -6.0);
        \coordinate (y) at (5.0, -6.0);
        \draw [thick, <->] (x) -- (y) node[below, pos=0.5] {$d_{\mathrm{pore}} =
        \SI{10}{\nano\meter}$};

    \end{tikzpicture}
    \caption{Sketch of the geometry used in the mean-field model. Details can be
    found in Ref.~\onlinecite{weik19b}.}
\label{fig:sketch origami}
\end{figure}
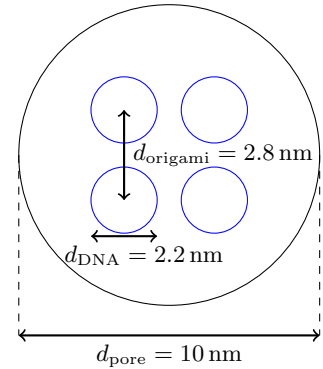

\section{Results}

\subsection{Model agreement across scales}\label{sec:results} 
\begin{figure}
	\includegraphics[width=\linewidth]{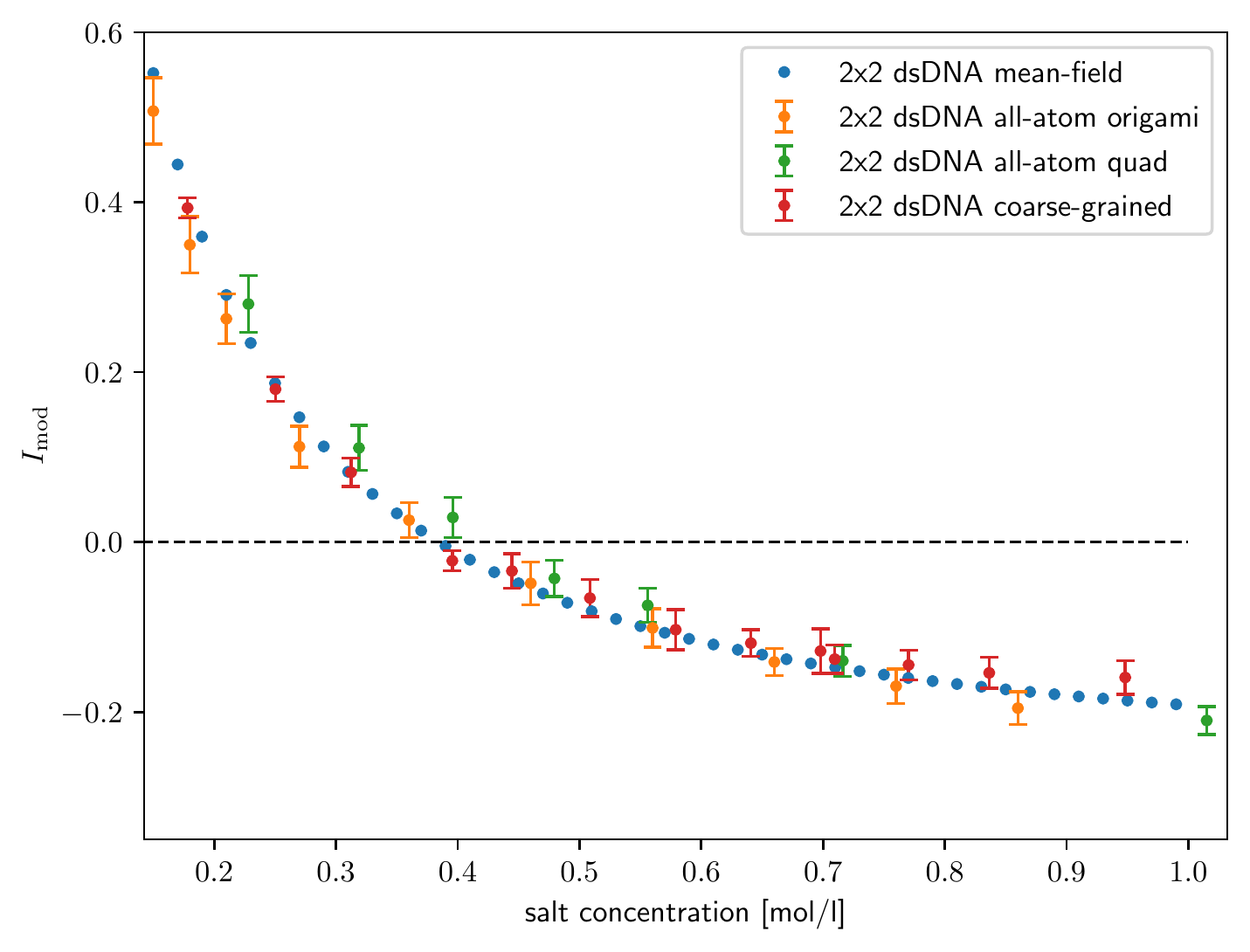}
	\caption{Relative change in ionic current through the nanopore for all four
		investigated models for a 2x2 dsDNA bundle. The results of all models
		agree within errors over the whole range of investigated salt
		concentrations.}
	\label{fig:2x2current}
\end{figure}

In the following, we will
compare the salt-dependent current modulation for the four aforementioned models
(all-atom origami, all-atom quad, coarse-grained and mean-field model). To get a
dimensionless number that we can directly compare across the models, we looked
at the relative change in ionic current $I_{\mathrm{mod}}=\left(I_{\mathrm{DNA}}
- I_{\mathrm{empty}}\right) / I_{\mathrm{empty}}$, where $I_{\mathrm{DNA}}$ is
the current of the pore with the DNA origami molecule inside and
$I_{\mathrm{empty}}$ the current of the empty pore with salt only. As shown in
Fig.~\ref{fig:2x2current}, the agreement for the current modulation is very good
among the models, taking into account the statistical errors for each model.
Especially with regard to the crossover salt concentration $\tilde{c}$ where the
relative change in current vanishes  ($I_\mathrm{mod}(\tilde{c})=0$), the
results of the four models agree within a small interval of the bulk salt
concentration. The fact that a quantitative agreement for these models has
already been observed for a single dsDNA molecule in
Refs.~\onlinecite{kesselheim14a,rau17a,weik19b} and the agreement we found in
this work for a larger dsDNA bundle structure, suggests that our models would
also show consistent results for larger bundle structures, \textit{e.g.} for a
system of 4x4 helices as investigated by Wang \textit{et al}. in
Ref.~\onlinecite{wang19a} (\textit{cf.} Fig.~\ref{fig:124}).

\begin{figure}
	\centering
	\includegraphics[width=\linewidth]{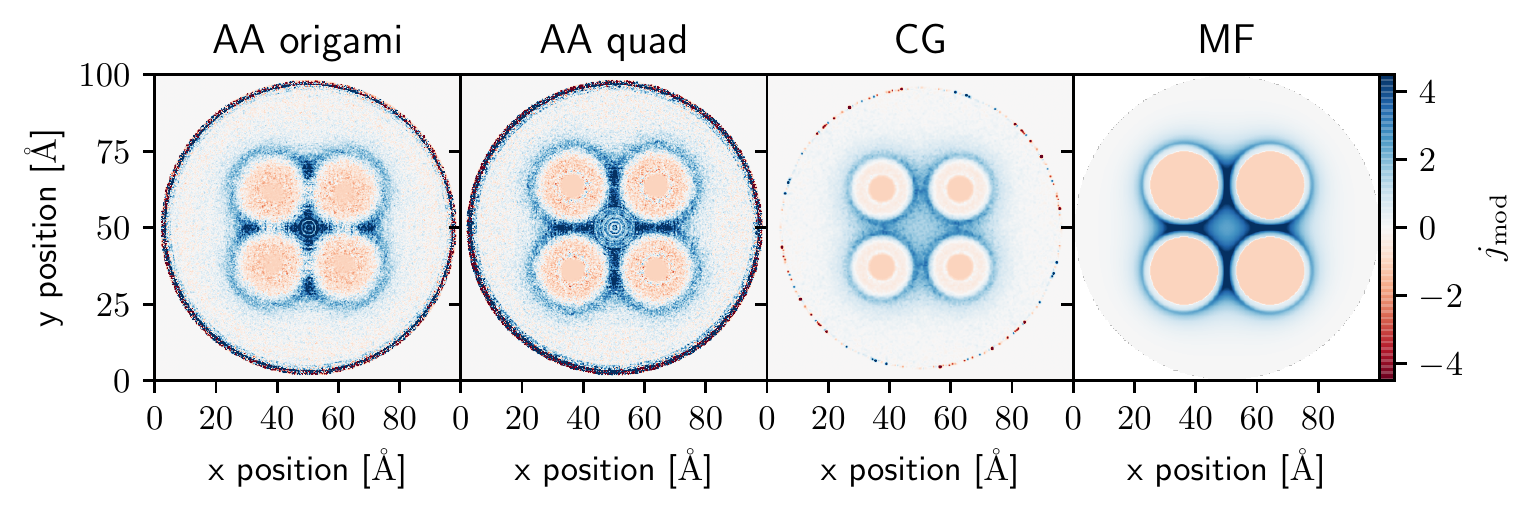}
	\caption{Relative current density modulation for the all-atom (left), the
		coarse-grained (mid) and the mean-field model (right). The non-vanishing
		modulation at the pore walls for the all-atom and coarse-grained models
		can be attributed to large current density fluctuations and poor
		statistics. The salt concentration for all models was approx.
		$\SI{0.18}{\mol\per\litre}$.}
	\label{fig:relcurrmodprof}
\end{figure}

In order to investigate the location dependent contribution to the current
modulation, the relative current density modulation across the pore
$j_{\mathrm{mod}} = \left(j_{\mathrm{DNA}} - j_{\mathrm{empty}}\right) /
j_{\mathrm{empty}}$ (naming convention as for $I_{\mathrm{mod}}$) is shown in
Fig.~\ref{fig:relcurrmodprof}. In case of the all-atom and the coarse-grained
model, the current density for the empty pore has been radially averaged to
reduce the noise. For the all-atom quad model, we additionally averaged the
current density data for the filled pore by rotating the data by $N\pi/2$ with
$N \in \{1, 2, 3\}$ around the pore center, leveraging the $D_4$ symmetry group
of the model's geometry. As expected, we observe a negative current density
modulation in the area where the DNA helices are located due to blockage and
friction with the ions. In the case of the all-atom origami model, the
footprints of the inter-helix connections are visible and neutralize the
modulation within a small area. For this model $j_\mathrm{mod}$ is larger in the
pore center than between two adjacent helices whereas in the case of the
all-atom quad model it is the other way around. The two coarser models
(coarse-grained and mean-field) without inter-helix connections both show a very
similar current density modulation profile compared to the all-atom quad model.

Since the current modulations for all investigated systems show a very
good agreement (\textit{cf.} Fig.~\ref{fig:2x2current}), the
differences in the current density modulation profiles (\textit{cf.}
Fig.~\ref{fig:relcurrmodprof}) have to have the following
characteristics: local differences between the models are canceled out
by complementary differences in other regions of the pore and these
compensating effects either do not depend on the salt concentration
\textit{or} depend on it in a way that preserves the ionic current's
dependency on the salt concentration.

\begin{figure}
	\centering
	\includegraphics[width=\linewidth]{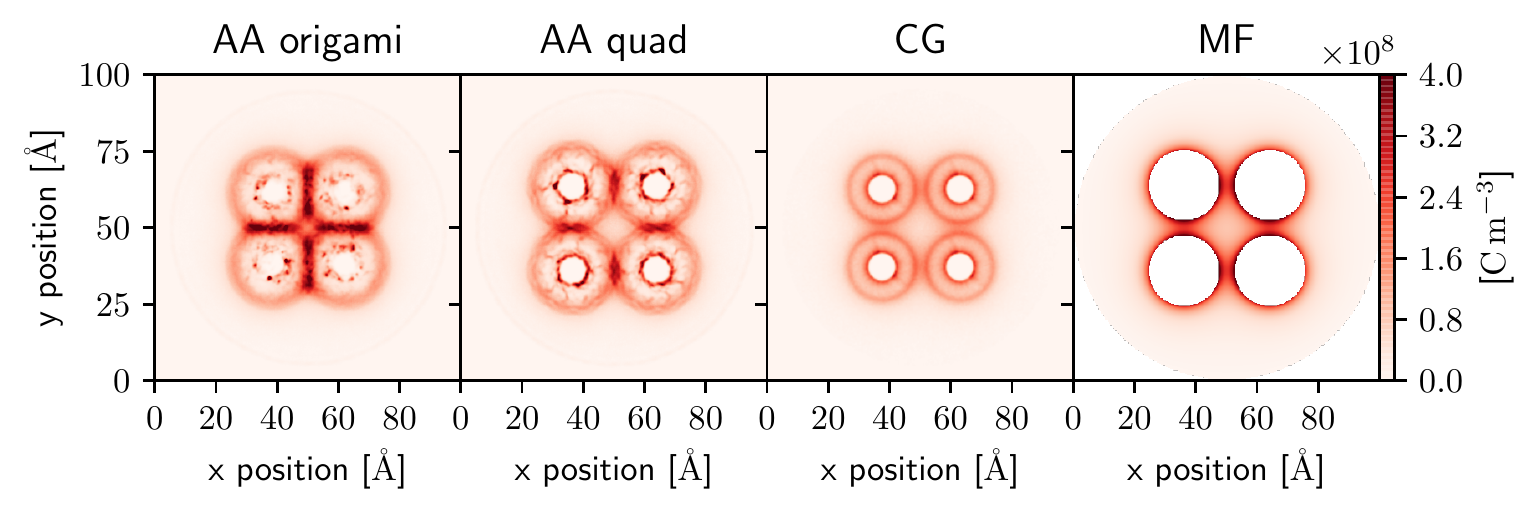}
	\caption{Charge density profiles for the all-atom origami, the all-atom
		quad, the coarse-grained and the mean-field model at approx.
		$\SI{0.18}{\mol\per\litre}$.}
	\label{fig:charge_density}
\end{figure}

To investigate the current density in greater detail, we additionally analyzed
the components that make up the current density, namely the charge density
profile and the ion velocity profiles. The charge density profiles shown in
Fig.~\ref{fig:charge_density} reveal that the all-atom origami system has a
slightly higher concentration of charges between adjacent helices but overall
the profiles for all four models look very similar. Thus, the charge density
profile does not explain the differences in the relative current density
modulation we observed for the all-atom origami model (\textit{cf.}
Fig.~\ref{fig:relcurrmodprof}).

The velocity profiles for the molecular dynamics systems (as described in
Sec.~\ref{sec:origami_model}, \ref{sec:quad_model}, \ref{sec:cg_model}) can be
directly computed from the trajectories of the ions. Ion velocities
\footnote{The impression of different pore diameters is an interpolation
artifact due to different bin sizes.} for the mean-field model have been
obtained via the following expression:
\begin{align}
	v_i(x,y)^z = \frac{j_i(x,y)^z}{c_i(x,y)} = \frac{u(x,y)^z+\mu_i ez_iE^z}{\alpha\mu\omega(x,y)+1}.
\end{align}
A detailed explanation of the notation can be found in Sec.~\ref{sec:mean_field_model}.

\begin{figure}
	\centering
	\includegraphics[width=\linewidth]{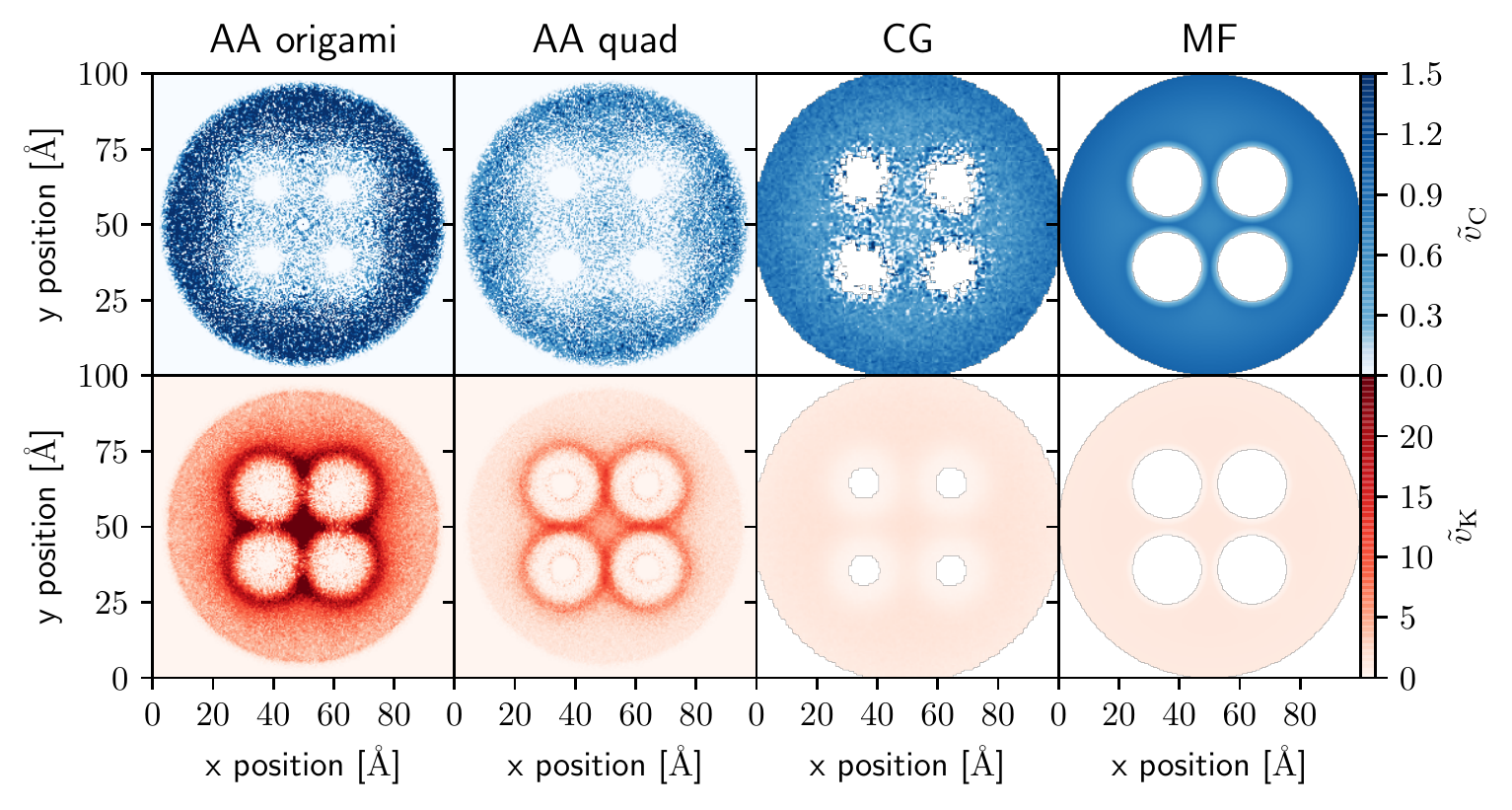}
	\caption{Normalized ion velocity profiles for the anions (top row) and the
		cations (bottom row) at a salt concentration of approx.
		$\SI{0.18}{\mol\per\litre}$. The velocities are normalized with the
		respective empty pore ion velocities in order to get directly comparable
		dimensionless values.}
	\label{fig:ion_velocity}
\end{figure}

All velocity profiles shown in Fig.~\ref{fig:ion_velocity} are normalized with
the respective average ion velocity in the empty pore, thus enabling the direct
comparison of the data for the different models. While the velocity profiles for
the anions do not show a significant deviation among the models, more prominent
differences are visible between the cation (counter-ion) velocity profiles of
the all-atom origami and all-atom quad model. Here, the cation velocity in the
origami model is larger at the pore center compared to the region between
adjacent helices. On the other hand, the opposite relation is observed for the
quad model. These differences in the velocity profiles together with the charge
density data shown in Fig.~\ref{fig:charge_density} thus explain the relative
current modulation differences between those two models.


\subsection{Comparison to experimental data}\label{sec:exp}
The group of Ulrich Keyser in Cambridge performed experiments in a
similar setting \cite{wang19a}. They conducted translocation
experiments using conical glass nanocapillary pores immersed in
solutions of KCl. These nanopores had a mean pore diameter of
\SI{14.2}{\nano\metre}. For comparison, we used \SI{10}{\nano\metre}
diameter pores in the simulations. However, as shown in a previous
publication \cite{weik19b} the pore size does not significantly
influence the current modulation. The DNA origami molecules of
Ref. \onlinecite{wang19a} have been designed and assembled of 4 or 16
parallel dsDNA double-helices connected by periodically repeating
crossover staple strands. Any possible occurrence of electroosmotic
flow due to the charged glass walls has been suppressed by adding a
tuned amount of polyethylene glycol. Wang \textit{et al}. reported a much
smaller crossover salt concentration
$\tilde{c}$(\SI{128}{\milli\mole\per\litre} for the 2x2 origami and
\SI{310}{\milli\mole\per\litre} for dsDNA) compared to similar
experimental setups where dsDNA translocation has been investigated
\cite{steinbock12a}.  Furthermore, they found a non-monotonic behavior
of $\tilde{c}$ with respect to the size of the analyte:
$\tilde{c}_{\text{dsDNA}} > \tilde{c}_{\text{2x2}} <
\tilde{c}_{\text{4x4}}$, \textit{i.e.} the crossover salt
concentration for 4x4 origami molecules is reportedly higher
(\SI{183}{\milli\mole\per\litre}) than for the smaller 2x2
origami. From all-atom simulations of an infinite pore system
\cite{kesselheim14a} with a dsDNA molecule it is known that the ionic
current is determined by two competing effects: (i) reduction in
current due to friction between ions and DNA helices, and (ii)
enhanced current due to additional mobile (counter-) ions. The
relative importance of these effects for the overall current
modulation depends on the bulk salt concentration of the electrolyte
and is non-trivial.

\begin{figure}
	\centering
	\includegraphics[width=\linewidth]{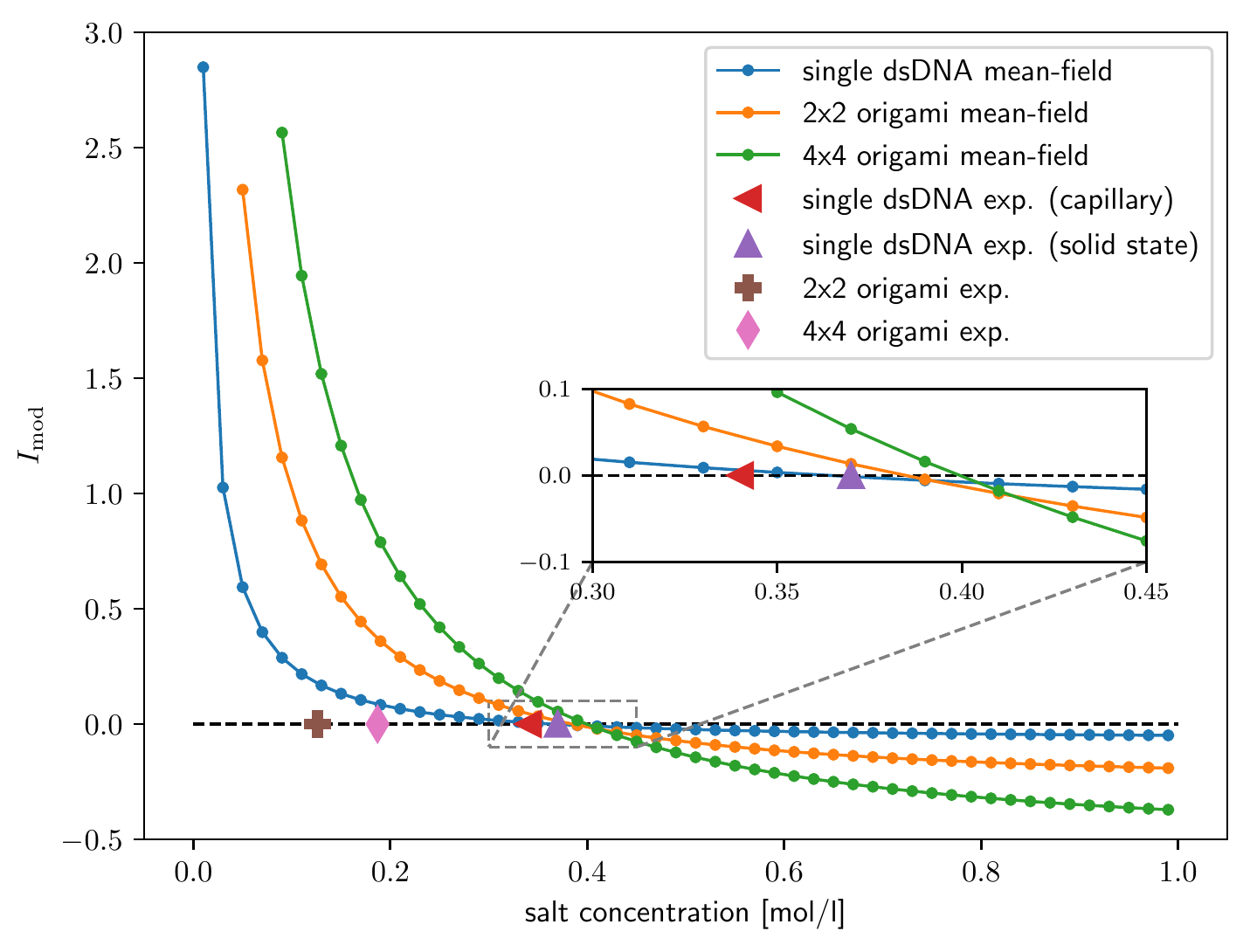}
	\caption{Comparison of the current modulation for different setups.
		Experimental data taken from Refs.~\onlinecite{smeets06a, wang19a}.}
	\label{fig:124}
\end{figure}

Since we now can safely assume that the frictional effects add up
linearly, we also investigated a 4x4 origami with our mean-field model.
An overview of the salt dependent current modulation for different
experimental systems and simulation models is shown in
Fig.~\ref{fig:124}. The data labeled as “mean-field” in the plot
legend refers to the respective model as described in
Sec.~\ref{sec:mean_field_model} with varying numbers of charged
cylinders representing the different numbers of DNA helices in the
model. The current modulation of the three mean-field systems shows a
slight monotonic trend towards higher crossover salt concentrations
for larger DNA structures (\textit{cf.} inset of Fig.~\ref{fig:124})
whereas the experimental results show a large drop of $\tilde{c}$ to
smaller salt concentrations from a dsDNA molecule to the 2x2
origami. Furthermore, the 4x4 origami shows an increased value for
$\tilde{c}$ compared to the 2x2 in the experiments of Wang et
al.~\cite{wang19a}. Thus, all of our simulation results show a
significant deviation for all origami systems despite being accurate
for the case of a single DNA molecule as has been shown in
Refs.~\onlinecite{kesselheim14a, weik16a, rau17a, weik19a}.

Comparing the experimental setup to our simulation models, we presume
deviations to result from one or combinations of the following
simplifications in our models: the pore entrance and the finite
molecule lengths are neglected, an alignment of the molecule's
symmetry axis with the pore axis is assumed, and the lateral position
of the DNA structure's symmetry axis is fixed to the pore center.

Regarding the finite pore and analyte effects, a possible deviation
might be expected since the 2x2 origami molecule and the 4x4 origami
molecule are folded from the same scaffold strand and therefore have
a different aspect ratios which may be related to the non-monotonic
effect of the molecule's apparent blocking area on the crossover salt
concentration. Moreover, Ref.~\onlinecite{wang19a} speculates about the
possibility of a diffusion limited current through the origami
structure.

Another possibility is that tilted conformations of the origami
molecules might occur that may lead to
a higher effective friction between ions and origami and the pore. As the
simulations show (\textit{cf.} Fig.~\ref{fig:relcurrmodprof}), the highest
current density is in the counter ion layer around the helices. If these layers
come closer to the pore walls, this might also lead to a reduction in pore
conductivity. However, due to the like-charge repulsion of the glass capillary
and the DNA molecule, we expect this effect to be minor.

The dependency of the pore conductivity on the position of a single dsDNA
molecule has already been investigated for the mean-field model in
Ref.~\onlinecite{weik19b}. No significant influence of the molecules position on
the pore conductivity had been found unless the gap between the DNA molecule and
the pore wall is smaller than the Debye length. Since the Debye length is
$\approx \frac{1}{10}$ pore diameter at the experimentally reported crossover
salt concentration, we do not expect this effect to be significant. Also, as our
current density profiles show that a non-negligible portion of the current stems
from ions inside the DNA structure which is not directly influenced by the pore
walls.

\section{Conclusion}
We presented a thorough investigation of four simulation models for the current
modulation of bundled DNA nanostructures. The coarse-grained and mean-field
models were parameterized only for a {\it single} dsDNA molecule. Although the level of
detail is ranging from the all-atom scale to the mean field scale, we observe a
very good agreement among the models with respect to the salt dependent current
modulation in a nanopore. This means that the frictional effects are additive
for the nanostructures which opens up the door to build arbitrarily large DNA
bundles from dsDNA. Spatially resolving the current density across the pore
revealed slight different ion mobilities at the center of the pore. While the
ion mobilities for the coarse-grained model and the mean-field model are very
similar, the all-atom origami model shows a higher ion mobility at the center of
the DNA nanostructure. The difference in the current density in the pore center,
however, is compensated by slight differences between adjoint helices in the
structure. In summary, our study shows that the current modulation for an
infinite pore system is robust against changes to molecular details.

Furthermore, we compared our results to recent experiments by Wang et
al.~\cite{wang19a} This comparison revealed a significant mismatch of the
crossover salt concentrations between our models and experimental results
(\textit{cf.} Fig.~\ref{fig:124}). Compared to a single dsDNA molecule, the
experimental results show a much lower ionic current if the bundled DNA
structure is in the pore \cite{wang19a}. Wang \textit{et al}. suggested this to
be a non-linear effect of the overlapping ion-DNA frictional interaction near
the bundled DNA structures. However, we do not observe such a drop in the
current density within the structures but a significant increase (compared to
the bulk) either between adjoint helices (all-atom quad, coarse-grained and
mean-field model) or at the pore center (all-atom DNA origami). The two other
possible suggestions for the current reduction in Ref.~\onlinecite{wang19a},
namely the idea of a diffusion limitation and end-effects in the pore, cannot be
investigated with the models presented here.

A possible way of further investigations could be a model with finite pore and
analyte. Such a model may be based on the coarse-grained or the mean-field model
we presented here and be extended by additional electrolyte cis- and
trans-reservoirs. Investigations along these lines are in progress.

\section{Acknowledgments}
We thank Jonas Landsgesell, Georg Rempfer, Alexander Schlaich and Ulrich Keyser
with co-workers for fruitful discussions. The work is funded by the Deutsche
Forschungsgemeinschaft (DFG, German Science Foundation) - Project Number
390740016 - EXC 2075. CH, KS, and FW also gratefully acknowledge funding by the
collaborative research center SFB 716.

\section{Data Availability}
The data that support the findings of this
study are available from the corresponding author upon reasonable request.

\bibliography{bibtex/icp.bib}

\clearpage
\section{Supporting Information}
\subsection{DNA origami structure}
\begin{figure}[h!]
	\begin{subfigure}{\linewidth}
		\centering
		\includegraphics[width=\linewidth]{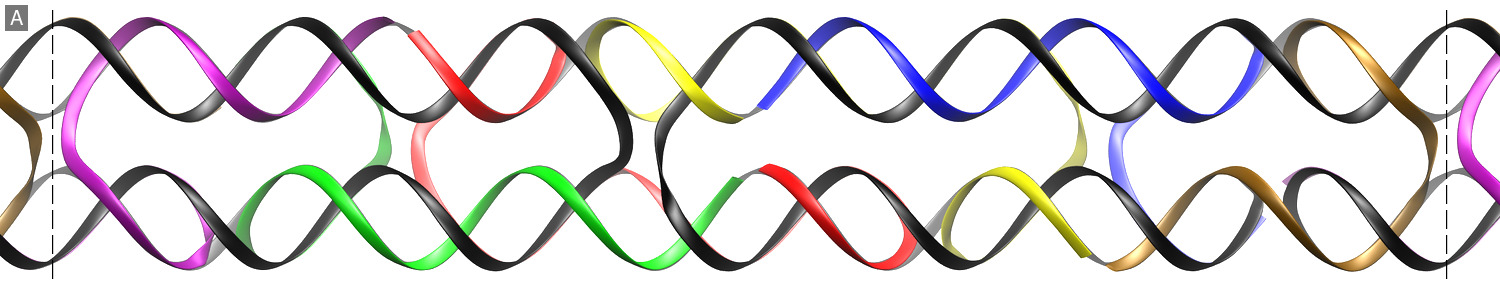}
	\end{subfigure}
	\begin{subfigure}{\linewidth}
		\centering
		\includegraphics[width=\linewidth]{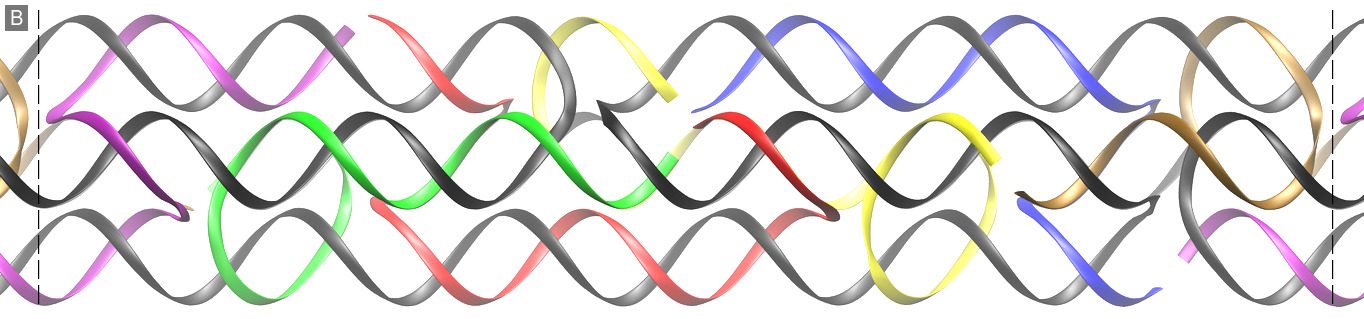}
	\end{subfigure}
	\caption{Close-up view of the unit cell from two different perspectives A
		and B related by a rotation of \SI{45}{\degree} around the main axis.
		The dashed lines represent the periodic boundaries. The scaffold strand
		is colored in black while the 6 linker strands are colored in blue, red,
		green, yellow, ocher and purple, respectively.}
	\label{fig:origami_unit}
\end{figure}

The DNA origami is a 4-helix bundle composed of a scaffold strand with 7250
nucleobases stabilized by 174 short staple strands. The DNA sequence of these
staple strands can be found in Ref.~\onlinecite{wang19a} Table S9. The quaternary
structure of the 4-helix bundle has translational symmetry along its main axis
with a length of \SI{217.6}{\angstrom}. An image of the unit cell can be found in
Fig.~\ref{fig:origami_unit}.

The structure used in this paper contains a segment of the scaffold strand
and 7 staple strands: full-length oligomers 9, 10, 67, 119, 150, and fragments
of oligomers 66 and 68.

\subsection{Interaction parameters of the coarse-grained model}
The mobility reduction observed in all-atom simulations \cite{kesselheim14a} is
modelled via a velocity dependent pair force between the backbone particles of
the coarse-grained model and the ions:
\begin{align}
	\vec{F}_{ij} = \begin{cases}
		-\zeta \left(1-\frac{r_{ij}}{r_\mathrm{c}}\right)^2\vec{v}_{ij}, & r_{ij} \leq r_\mathrm{c}\\
		0, & \mathrm{else},
	\end{cases}
\end{align}
where $\zeta$ is the friction parameter, $r_c$ the cutoff distance, $r_{ij}$ the
distance between the particles $i$ and $j$, $\vec{v}_{ij}$ is the relative
velocity. We chose the two free parameters $r_c=\SI{11}{\angstrom}$ and
$\zeta=6.25$ such that the velocity profiles between the coarse-grained model
and the results from all-atom simulations matched reasonably well.

Furthermore, the non-bonded Lennard-Jones interaction potential has been used to model
the excluded volume between particles:
\begin{equation}
	\begin{split}
		U(r_{ij}) =
		\begin{cases}
			4 \epsilon \left[ \left(\frac{\sigma}{r_{ij}-r_\mathrm{off}}\right)^{12}
				- \left(\frac{\sigma}{r_{ij}-r_\mathrm{off}}\right)^6+c\right]
			 &                \\ \text{ for } r_\mathrm{off} < r_{ij} < r_\mathrm{c}+r_\mathrm{off} \\
			0 ,
			 & \mathrm{else},
		\end{cases}
	\end{split}
	\label{eq:LJ}
\end{equation}
where $\sigma$ is the particle diameter, $r_{ij}$ the distance between the
particles, $r_{\mathrm{off}}$ a distance offset, $r_\mathrm{c}$ the cutoff
distance and $c$ an energy shift. In the following we list the interaction
parameters for the different particle pairs of the coarse-grained model.
\begin{table}[h]
	\centering
	\caption{Summary of pair interaction parameters.}
	\begin{tabular}{c|c|c|c|c}
		interaction pair            & $\sigma [\si{\angstrom}]$   & $\epsilon [\kb T]$ &
		$r_{\mathrm{cut}} [\sigma]$ & $r_{\mathrm{off}} [\sigma]$                                                 \\
		\hline
		ion-ion                     & 4.25                        & 1                  & $2^{\frac{1}{6}}$ & 0.0  \\
		ion-pore                    & 4.25                        & 1                  & $2^{\frac{1}{6}}$ & 0.0  \\
		ion-backbone                & 4.25                        & 1                  & $2^{\frac{1}{6}}$ & 0.0  \\
		ion-basepair                & 4.25                        & 1                  & $2^{\frac{1}{6}}$ & 0.18
	\end{tabular}
\end{table}
The parameter $c$ in Eq.~\eqref{eq:LJ} is chosen such that the interaction
potential is zero at the chosen cutoff distance. Since the DNA is fixed in the
pore center there is no interaction defined between DNA beads and the
pore.
\end{document}